\documentclass[aps,prl,reprint,showpacs,showkeys,superscriptaddress,floatfix,twocolumn]{revtex4-2}
\include{header}
\usepackage{graphicx}
\usepackage{dcolumn}
\usepackage{bm}
\usepackage[mathlines]{lineno}
\usepackage{multirow}
\usepackage{natbib}
\usepackage{amsmath}
\usepackage{times}
\usepackage{xcolor}
\usepackage{soul}
\usepackage[normalem]{ulem}
\usepackage{textcomp,mathcomp}
\usepackage{bbold}
\usepackage{mathtools}
\usepackage{braket}
\usepackage{soul}
\usepackage{xr} 
\usepackage[colorlinks=true,linkcolor=magenta,citecolor=magenta,hypertexnames=false]{hyperref}

\def\ba#1\ea{\begin{align}#1\end{align}}
\def\bg#1\eg{\begin{gather}#1\end{gather}}
\def\bpm{\begin{pmatrix}}
	\def\epm{\end{pmatrix}}

\newcommand{\ourtitle}{Inhomogeneous dynamic state in the double trillium lattice antiferromagnet KBaFe$_2$(PO$_4$)$_3$}

\allowdisplaybreaks

\begin{document}
\title{\textbf{\ourtitle}}
	
\author{Sebin J. Sebastian }
\affiliation{School of Physics, Indian Institute of Science Education and Research Thiruvananthapuram-695551, India}
\affiliation{Ames National Laboratory, U.S. DOE, Iowa State University, Ames, IA 50011, USA}
\author{Shams Sohel Islam}
\email{shams.islam@psi.ch}
\affiliation{PSI Center for Neutron and Muon Sciences CNM, 5232 Villigen PSI, Switzerland}
\author{R. Kolay}
\author{S. Mohanty}
\affiliation{School of Physics, Indian Institute of Science Education and Research Thiruvananthapuram-695551, India}
\author{Q.-P. Ding }
\affiliation{Ames National Laboratory, U.S. DOE, Iowa State University, Ames, IA 50011, USA}
\author{Y. Skourski}
\affiliation{High Magnetic Field Laboratory (HLD-EMFL), Helmholtz-Zentrum Dresden-Rossendorf, 01328 Dresden, Germany}
\author{J. Sichelschmidt}
\affiliation{Max-Planck-Institut f{\"u}r Chemische Physik fester Stoffe, 01187 Dresden, Germany}
\author{M. Baenitz}
\affiliation{Max-Planck-Institut f{\"u}r Chemische Physik fester Stoffe, 01187 Dresden, Germany}
\author{Jonas A. Krieger}
\affiliation{PSI Center for Neutron and Muon Sciences CNM, 5232 Villigen PSI, Switzerland}
\author{T. J. Hicken}
\affiliation{PSI Center for Neutron and Muon Sciences CNM, 5232 Villigen PSI, Switzerland}
\author{H. Luetkens}
\affiliation{PSI Center for Neutron and Muon Sciences CNM, 5232 Villigen PSI, Switzerland}
\author{A. A. Tsirlin}
\affiliation{Felix Bloch Institute for Solid-State Physics, Leipzig University, 04103 Leipzig, Germany}
\author{Y. Furukawa}
\affiliation{Ames National Laboratory, U.S. DOE, Iowa State University, Ames, IA 50011, USA}
\affiliation{Department of Physics and Astronomy, Iowa State University, Ames, IA 50011, USA}
\author{R. Nath}
\email{rameshchandra.nath@gmail.com}
\affiliation{School of Physics, Indian Institute of Science Education and Research Thiruvananthapuram-695551, India}

\date{\today}
\begin{abstract}
The three-dimensional (3D) magnet KBaFe$_2$(PO$_4$)$_3$ hosts a double-trillium lattice of Fe$^{3+}$ (spin, $S=5/2$) ions offering a prototypical platform to study the frustration induced effects in 3D. Through magnetization, specific heat,$^{31}$P nuclear magnetic resonance (NMR), and muon spin relaxation ($\mu$SR) experiments, supported by first‑principles calculations, we uncover an unconventional ground state. Despite strong antiferromagnetic interactions with a large Curie-Weiss temperature $\theta_{\rm{CW}}= -70(2)$~K, no magnetic long‑range order is observed down to 30\,mK. Below $T^{*}\simeq3.5$~K, the NMR linewidth becomes nearly field‑independent and the spin-spin relaxation rate $1/T_2$ saturates, accompanied by an inhomogeneous distribution of transverse nuclear magnetization $M_{xy}$. The latter indicates the emergence of short‑range dynamical correlations, which was further corroborated by a robust and field‑insensitive broad maximum in specific heat. 
In $\mu$SR, we detect neither a static internal field nor spin-freezing; instead the relaxation remains dynamic and is best described by two coexisting dynamic relaxation channels: a dominant fast (sporadic) channel and a slower Markovian component. Their differing weights and fluctuation rates suggest microscopic inhomogeneity in spin dynamics. Altogether, KBaFe$_2$(PO$_4$)$_3$ exemplifies a rare high‑spin stoichiometric 3D antiferromagnet that evades ordering and instead fosters a mosaic of spin dynamics driven by strong geometric frustration intrinsic to the trillium lattice.
\end{abstract}

\maketitle
\let\oldaddcontentsline\addcontentsline
\renewcommand{\addcontentsline}[3]{}

Frustrated antiferromagnets (AFM) are at the forefront of condensed matter research, since frustration can destabilize conventional magnetic long-range order (LRO), leading to degenerate states and hence, numerous collective phenomena. In particular, quantum spin liquids (QSLs) represent a remarkable manifestation of magnetic frustration; a disordered yet highly entangled magnetic ground state characterized by fractionalized excitations, persistent spin dynamics without any symmetry breaking even at $T=0$~K, and $\text{U(1)}$, $\mathbb{Z}_2$, or $\text{SU(2)}$ gauge fields~\cite{Balents199,Savary016502,Wang14}. While two-dimensional (2D) frustrated magnets have historically dominated the search for QSLs following Anderson’s resonating valence bond (RVB) hypothesis~\cite{Anderson153}, there is a growing perception that three-dimensional (3D) frustrated lattices also offer ample opportunities for realizing similar rich physics.

While the hyperkagome~\cite{Okamoto137207,Zhou197201,Lawler227201,Khuntia107203,Hong256701,Chillal2348} and pyrochlore lattices~\cite{Zhang167203,Plumb54,Lee47,Gao1052} are two well-studied geometries in 3D, the trillium lattice is another intriguing addition to this category.
Here, a chiral network of corner-sharing triangles with three-fold connectivity induces magnetic frustration~\cite{Hopkinson224441}. The chiral nature of the spin network plays a key role in stabilizing magnetic skyrmions and chiral phonons, and also lead to topological Hall effect in itinerant magnets~\cite{Muhlbauer915,Neubauer186602}.
Recently, compounds of the langbeinite family featuring bipartite variant \emph{double-trillium} geometry have received significant attention. For instance, the double trillium AFMs K$_2$Ni$_2$(SO$_4$)$_3$ (Ni$^{2+}$; $S=1$) and K$_2$Co$_2$(SO$_4$)$_3$ (Co$^{2+}$; $J_{\rm eff}=1/2$)  feature a field-induced QSL-like state and are projected to be in the vicinity of the quantum critical regime~\cite{Ivica157204,Magar2025}. Similarly, KBaCr$_2$(PO$_4$)$_3$ (Cr$^{3+}$; $S=3/2$) demonstrates field-induced magnetic transitions~\cite{Kolay224405}, while KSrFe$_2$(PO$_4$)$_3$ (Fe$^{3+}$; $S=5/2$) is proposed to be a spin-liquid (SL) candidate~\cite{Boya101103}. Our recent $^{31}$P nuclear magnetic resonance (NMR) studies on KSrFe$_2$(PO$_4$)$_3$ revealed short-range spin freezing in a highly dynamic ground state~\cite{Sebin2025}. On the other hand, the metal-organic framework Na[Mn(HCOO)$_3$] (Mn$^{2+}$; $S=5/2$) that features trillium geometry exhibits an exotic 2-\textbf{k} magnetic ordering accompanied by a pseudo 1/3 magnetization plateau~\cite{Bulled177201}. These examples underscore the potential of trillium lattices to exhibit interesting low-temperature phenomena, which are often influenced by the spin value, disorder, magnetic anisotropy, and hierarchy in magnetic interactions.

\begin{figure}
\centering
\includegraphics[width=\linewidth]{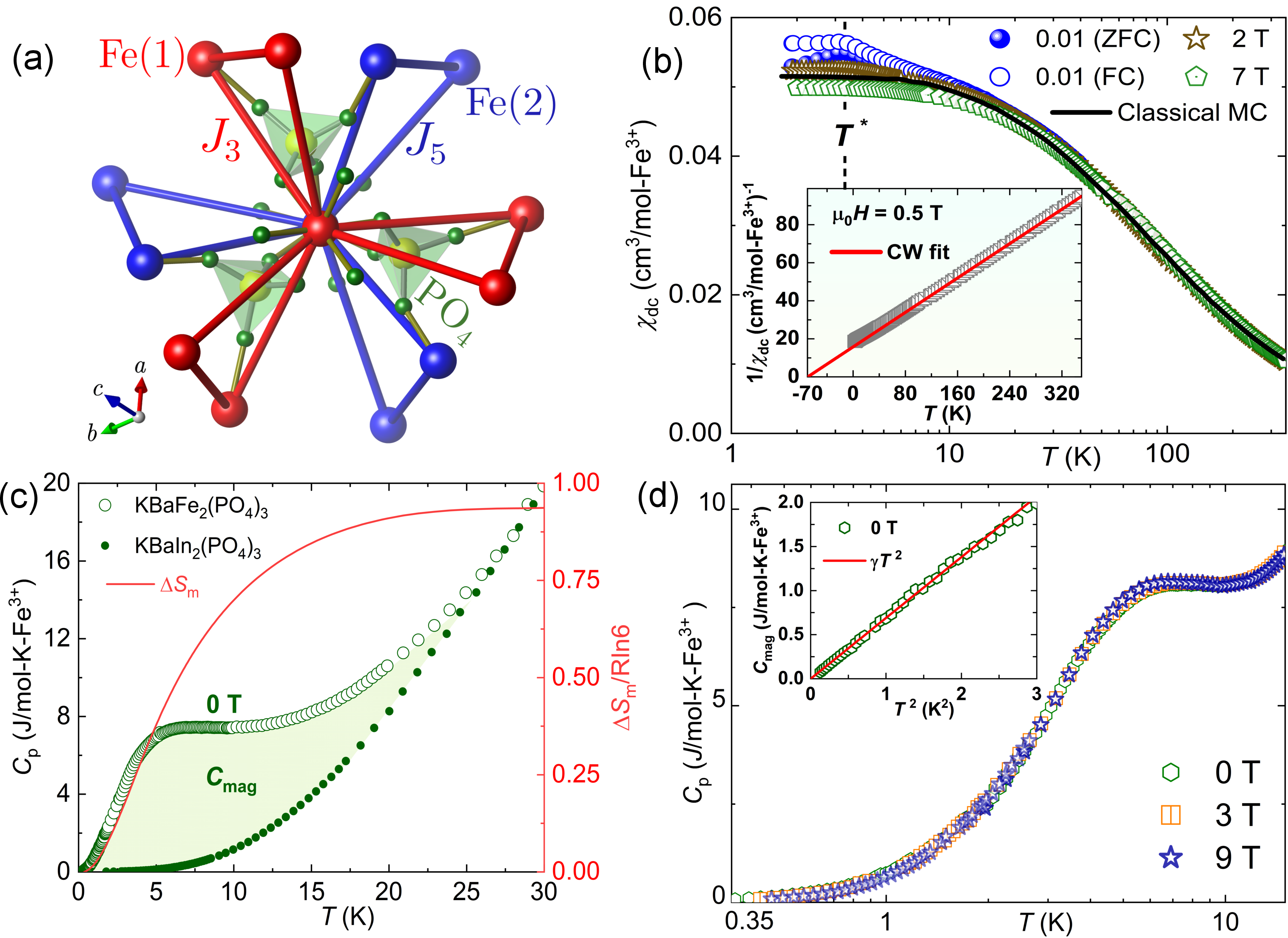}
\caption{(a) A coupled trillium unit of Fe$^{3+}$ ions, connected via PO$_4$ tetrahedra. (b) $\chi_{\rm dc}$ vs $T$ measured under ZFC and FC conditions in $\mu_0 H = 0.01$~T and in other applied fields. The solid line represents the classical Monte-Carlo simulation using the optimized exchange parameters from Table.~\ref{tab:exchange}. Inset: CW fit to $\chi_{\rm dc}(T)$. (c) Left $y$-axis: $C_{\rm p}$ vs $T$ for KBFPO and its non-magnetic analogue KBaIn$_2$(PO$_4$)$_3$ measured in zero-field and right $y$-axis: magnetic entropy $\Delta S_{\rm m}(T)$ calculated from $C_{\rm mag}(T)$. (d) $C_{\rm p}(T)$ measured in different applied fields. Inset: $C_{\rm mag}$ vs $T^2$ plot fitted with a straight line at low-$T$s.}
\label{Fig1}
\end{figure}

In this Letter, we present an in-depth investigation of spin dynamics and static magnetic properties of KBaFe$_2$(PO$_4$)$_3$ (KBFPO) employing local probes like $^{31}$P NMR and muon spin resonance ($\mu$SR), dc and ac susceptibility, and specific heat measurements, complemented by density-functional theory (DFT) calculations. Being a member of the langbeinite family, KBFPO crystallizes in the cubic space group $P2_13$, where Fe$^{3+}$ ($S = 5/2$) ions are embedded in a double-trillium lattice [see Fig.~\ref{Fig1}(a)]. Owing to its highly frustrated geometry, KBFPO evades LRO, with AFM fluctuations prevailing down to 30~mK. Our local probes strongly suggest inhomogeneous spin dynamics of the fluctuating spins; a notable feature observed for the first time in a double trillium lattice AFM. The details about polycrystalline sample synthesis and characterization, as well as experimental protocols for dc and ac magnetic susceptibility, specific heat, $^{31}$P-NMR, and $\mu$SR experiments, along with the DFT calculations are given in the Supplementary Material (SM)~\cite{SM}.

Figure~\ref{Fig1}(b) depicts the temperature-dependent dc magnetic susceptibility ($\chi_{\rm dc}$) measured in different applied fields. In a low magnetic field of 0.01~T, a weak bifurcation emerges between $\chi_{\rm dc}$ measured under zero-field-cooled (ZFC) and field-cooled (FC) conditions near the characteristic temperature, $T^{*} \simeq 3.5$~K, which disappears completely in magnetic fields exceeding 1~T. A Curie-Weiss (CW) fit [inset of Fig~\ref{Fig1}(b)] to $\chi_{\rm dc}(T)$ at high temperatures ($T > 100$~K) returns an effective magnetic moment, $\mu_{\rm eff} = 5.93(2)\mu_{\rm B}$ (consistent with the spin-only value of $\mu_{\rm eff} = 5.91\mu_{\rm B}$ for $S = 5/2$ and $g=2$) and a CW temperature $\theta_{\rm CW} = -70(2)$~K. Further, it is observed that $\chi_{\rm dc}$ at low temperatures shows the tendency towards saturation for $\mu_0 H \geq 1$~T, which does not support a conventional AFM LRO near $T^{*}$. Complementing $\chi_{\rm dc}(T)$, the specific heat [$C_{\rm p}(T)$] reveals no signature of magnetic LRO down to 0.35~K, but exhibits a pronounced broad maximum at around 7~K [Fig.~\ref{Fig1}(c)]. This broad maximum is found to be field-independent [Fig.~\ref{Fig1}(d)], evocative of a robust AFM short-range order, similar to other 3D SL candidates, Na$_4$Ir$_3$O$_8$~\cite{Okamoto137207} and PbCuTe$_2$O$_6$~\cite{Koteswararao035141}. Typically, for 3D frustrated and SL systems, the broad maximum in $C_{\rm p}(T)$ is expected to appear roughly in the range of $T/\theta_{\rm CW}\simeq 0.02-0.15$~\cite{Koteswararao035141}. Indeed, for KBFPO, this ratio is close to 0.055, which falls well within the expected range, indicating the proximity of KBFPO to the SL state. The phonon contribution [$C_{\rm ph}(T)$] to the total specific heat is subtracted using an isostructural non-magnetic analog KBaIn$_2$(PO$_4$)$_3$. After subtraction, the magnetic specific heat [$C_{\rm mag}(T)$] below 1~K is fitted by $C_{\rm mag}(T) =\gamma T^{\alpha}$, which shows a quadratic behaviour ($\alpha=2$) [inset of Fig.~\ref{Fig1}(d)] with a Sommerfeld coefficient $\gamma \simeq 0.69$~J/mol-K$^2$. Given the 3D nature of the material, this reduced exponent excludes the possibility of a 3D AFM LRO and indicates unconventional magnetic excitations, as found in several highly frustrated magnets with SL state~\cite{Nakatsuji1697,Ivica157204,Plumb54}. Similarly, though the real part of the ac susceptibility ($\chi^{\prime}_{\rm ac}$) features a weak anomaly near $T^{*}$, this anomaly is frequency-independent, inconsistent with the spin-glass scenario~\cite{SM}. 

\begin{table}
\caption{
\label{tab:exchange}
Exchange couplings $J_i$ (in K) in KBFPO, calculated in the present work using DFT, and optimized to achieve the best fit of the experimental magnetic susceptibility (Fig.~\ref{Fig1}). The labels of $J_1-J_5$ follow Ref.~\cite{Gonzalez7191}. }
\begin{ruledtabular}
\begin{tabular}{cccccc}
                           & $J_1$ & $J_2$ & $J_3$ & $J_4$ & $J_5$ \smallskip\\
$d_{\rm Fe-Fe}$ (in\,\r A) & 4.562 & 4.939 & 6.080 & 6.039 & 6.146 \smallskip\\
DFT                        &  1.7  &  1.3  &  2.5  &  4.3  &  3.5  \\
Optimized                  &  1.5  &   0   &  1.6  &  2.7  &  3.9  \\
\end{tabular}
\end{ruledtabular}
\end{table}

Using DFT, we find that all five couplings of the double-trillium lattice in KBFPO are AFM and, therefore, highly frustrated. We further refine the $J_1-J_5$ values by a direct comparison with the experimental $\chi_{\rm dc}(T)$ (Fig.~\ref{Fig1}). Similar to K$_2$Ni$_2$(SO$_4)_3$~\cite{Ivica157204}, the two dominant couplings in KBFPO are $J_4$ and $J_5$, but finite $J_1$ and $J_3$ are also essential to describe $\chi_{\rm dc}(T)$ at low temperatures. Given the spin-5/2 nature of Fe$^{3+}$, the $T^*$ is at least one order of magnitude smaller than the energy scale of the exchange couplings. 

\begin{figure*}
\centering
\includegraphics[width=\textwidth]{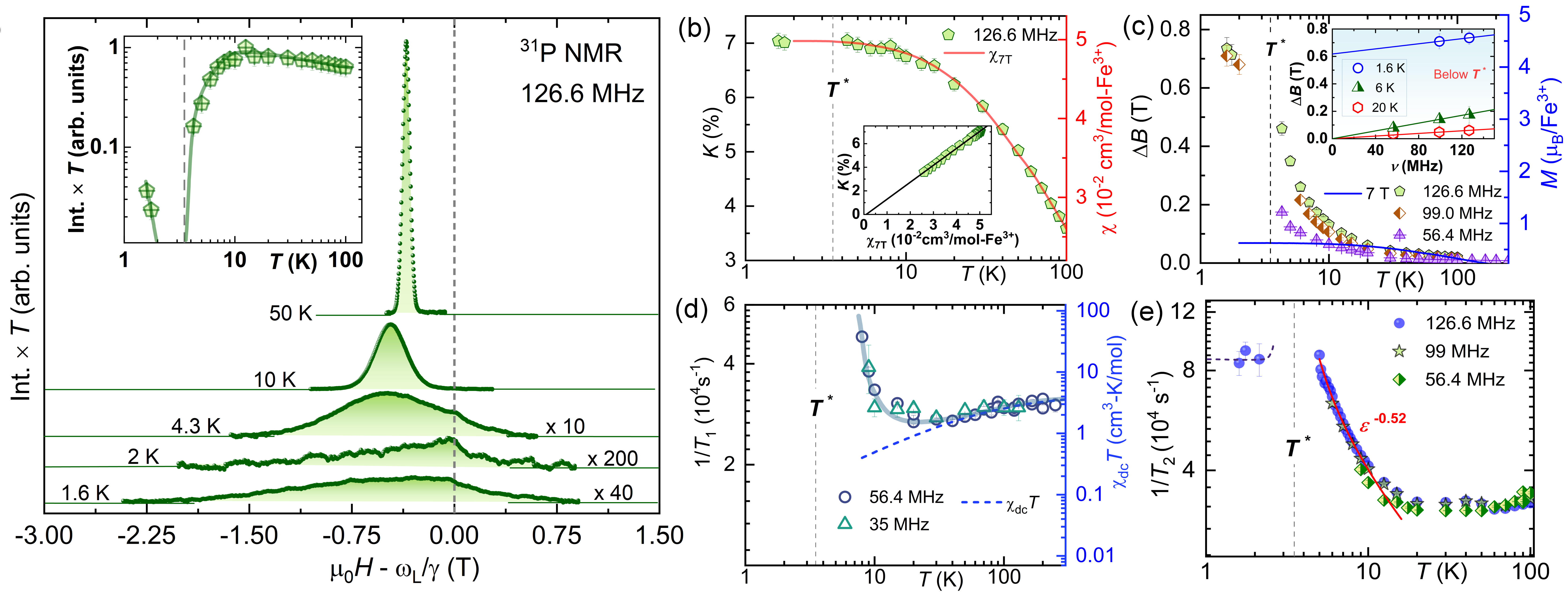}
\caption{(a) Temperature dependent field-sweep $^{31}$P NMR spectra measured at 126.6~MHz. The vertical dashed line marks the Larmor field. Inset: temperature-dependent signal intensity multiplied by temperature with the $T_2$ corrections. (b) Temperature dependence of the $^{31}$P NMR shift ($\mathcal{K}$) measured at 126.6 MHz, overlaid with $\chi_{\rm dc}$ measured at 7~T. Inset: $\mathcal{K}$ vs $\chi_{\rm dc}$. (c) Full width at half maximum ($\Delta B$) of the $^{31}$P NMR line as a function of temperature, measured at three different frequencies. The solid line represents the magnetization ($M$) vs $T$ plot for $\mu_0 H = 7$~T. Inset: $\Delta B$ as a function of the NMR frequency. (d) Temperature dependence of $1/T_1$ at different frequencies, together with the temperature dependence of $\chi_{\rm dc}T$ measured at 2 T. (e) $1/T_2$ as a function of temperature measured at various frequencies. The red solid line represents the fit discussed in the text, with $T^{*} = 3.5$ K.}
\label{Fig2}
\end{figure*}
To probe the ground state of KBFPO, we performed a comprehensive $^{31}$P NMR investigation and the important results are summarized in Fig.~\ref{Fig2}. 
As shown in Fig.~\ref{Fig2}(a), the NMR spectra exhibit a single and nearly Gaussian lineshape at high temperatures, consistent with the unique crystallographic P-site. Upon cooling, the NMR line broadens and shifts to lower fields. Below $T \sim 10$~K, the NMR signal intensity decreases rapidly and becomes undetectable by the NMR spectrometer while approaching $T^*$ [inset of Fig.~\ref{Fig2}(a)]. This wipe-out effect results from the rapid shortening of spin-spin relaxation time ($T_2$), pushing the spin echo below the instrumental resolution. For $T \lesssim 2$~K, we observed a partial recovery of the signal, though only about $\sim 3$\% of the intensity (as compared to high temperatures) is recovered at 1.6~K. This behavior draws parallel with that observed in other highly frustrated magnets featuring unusual spin dynamics~\cite{Takeya054429,Olariu167203}. The NMR line retains a Gaussian shape across $T^{*}$, inconsistent with a rectangular pattern expected for a conventional AFM ordering.

The extracted NMR shift $\mathcal{K}(T)$ from the spectral peak positions [Fig.~\ref{Fig2}(b)] follows $\chi_{\rm dc}(T)$ very well, confirming the intrinsic origin of the NMR response. A linear $\mathcal{K}$–$\chi$ plot [inset of Fig.~\ref{Fig2}(b)] yields a hyperfine coupling constant $\mathcal{A}_{\rm hf} = 0.80(1)$~T/$\mu_{\rm B}$ between the $^{31}$P nucleus and the Fe$^{3+}$ spins and a temperature-independent orbital shift $\mathcal{K}_0 = -0.14(1)\%$. The saturation of $\mathcal{K}(T)$ to a finite value at low temperatures supports the formation of a gapless state in KBFPO. Moreover, below 10~K, we see a significant increase in the NMR linewidth ($\Delta B$) for all measured frequencies [Fig.~\ref{Fig2}(c)]. As presented in the inset of Fig.~\ref{Fig2}(c), for $T>T^{*}$, $\Delta B$ scales linearly with $\nu$ (or, $H$) as expected for the paramagnetic fluctuations. However, below $T^{*}$ (i.e., at 1.6~K), the linewidth remains almost field-independent with a residual linewidth [$\Delta B(0) = 0.6(1)$~T] for $\nu\rightarrow0$.
Thus, the NMR spectral analysis reveals 
the existence of a finite internal field well below $T^{*}$.

Figure~\ref{Fig2}(d) presents the spin-lattice relaxation rate ($1/T_1$) measured at different frequencies. In general, $1/T_1$ is given by $1/T_1 = (2\gamma^{2}_{\rm N}k_{\rm B}T/N^{2}_{\rm A}) \sum_{\vec{q}} |\mathcal{A}(\vec{q})|^2 \chi''(\vec{q},\omega_{\rm N})/\omega_{\rm N}$, where $\mathcal{A}(\vec{q})$ is the hyperfine form factor and $\chi''(\vec{q},\omega_{\rm N})$ is the imaginary part of the dynamic susceptibility at the Larmor frequency $\omega_{\rm N}$~\cite{Moriya516,Nath214430}. At high temperatures, $1/T_1$ scales with $\chi_{\rm dc}T$, indicating that the spin-lattice relaxation is dominated by paramagnetic fluctuations. This also reflects that $\sum_{\vec{q}} |\mathcal{A}(\vec{q})|^2 \chi''(\vec{q},\omega_{\rm N})$ follows the same temperature dependence as the static uniform susceptibility $\chi'(0,0)$. Below $\sim 20$~K, $1/T_1$ exhibits a clear enhancement beyond the trend expected from $\chi_{\rm dc}T$, signaling the growth of dynamic spin correlations with wave vector $q\neq 0$. This enhancement is attributed to the slowing down of AFM fluctuations on approaching $T^{*}$. Notably, the increase in $1/T_1$ occurs over a broad temperature range, in contrast to a sharp divergence typically expected for a conventional AFM LRO~\cite{Devi015803}, further underscoring the unconventional nature of spin dynamics in KBFPO. 

Since $T_1$ becomes too short to be measured below 8~K, we employed spin-spin relaxation rate ($1/T_2$) to probe the low-temperature dynamics, which was measurable very close to $T^{*}$. As shown in Fig.~\ref{Fig2}(e), $1/T_2$ vs $T$ closely mirrors that of $1/T_1$, with a broad minimum at around 20~K and a sharp upturn below 10~K. It saturates to a large and nearly constant value below $T^{*}$, reflecting the presence of persistent spin dynamics. A power-law fit in the critical region ($T^{*}\leq T\leq 2T^{*}$) using $1/T_2 \propto \epsilon^{-\beta}$ (where, $\epsilon = T/T^* - 1$) yields a critical exponent $\beta=0.52(3)$, larger than the typical value predicted for a conventional 3D AFM order ($\sim 0.3$)~\cite{Devi015803}.
Also, below $T^{*}$ the transverse nuclear magnetization ($M_{xy}$) follows a double-exponential behaviour in contrast to the single exponential behaviour above $T^{*}$~\cite{SM}. Together, these observations, namely (\textit{i}) the absence of conventional AFM LRO, (\textit{ii}) the emergence of multiple relaxation channels in $M_{xy}$ below $T^*$, and (\textit{iii}) the enhanced critical behavior of $1/T_2$ with an unusual value of $\beta$, imply an inhomogeneous and dynamically fluctuating ground state in KBFPO.

\begin{figure*}
\includegraphics[width = 0.85\textwidth]{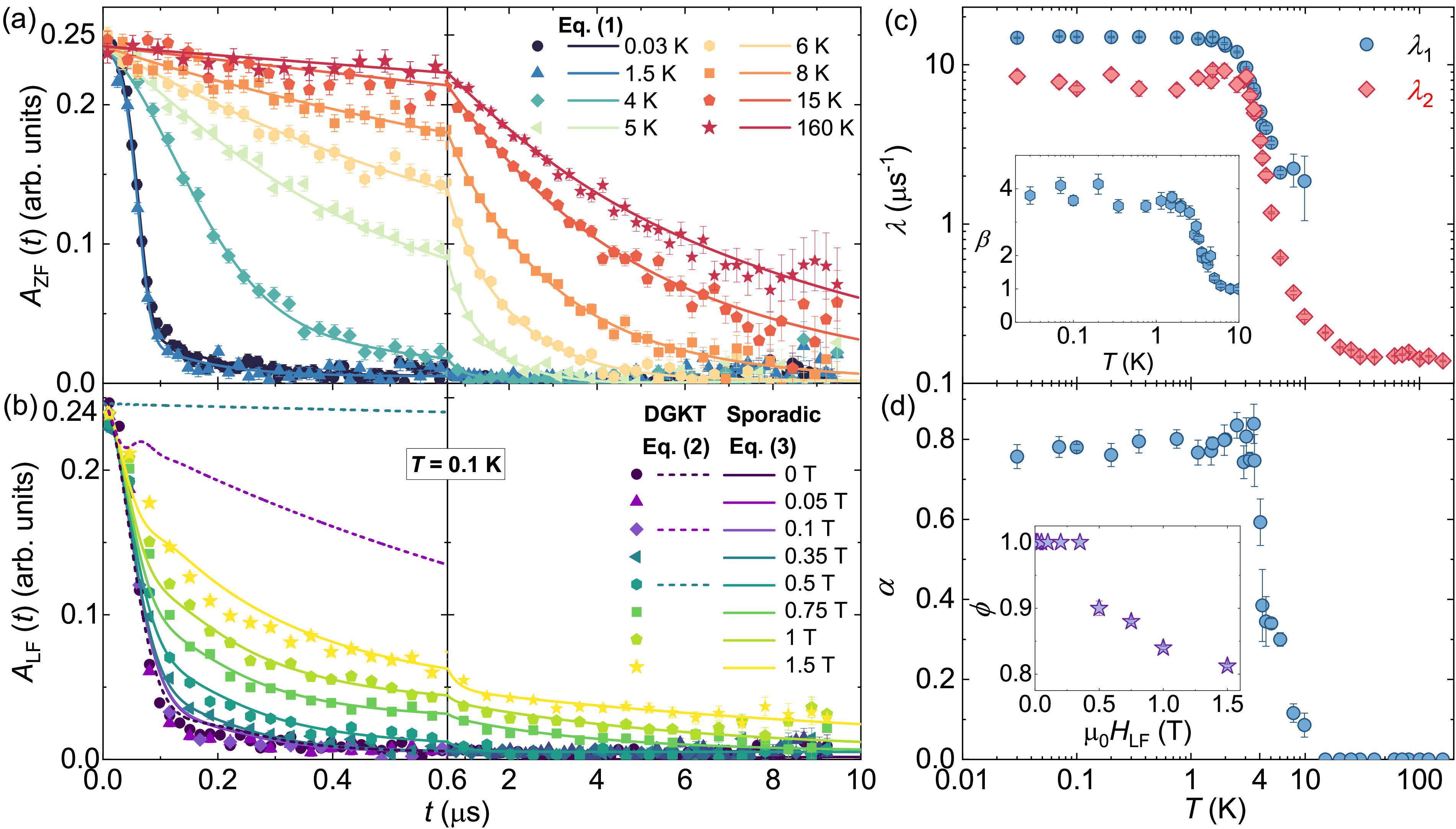}
\caption{\label{Fig3}
(a) ZF muon spin asymmetry [$A_{\rm ZF}(t)$] at different temperatures. Left panel: magnified short-time behavior ($t \le 0.6\,\mu$s) and right panel: long-time tail ($t \ge 0.6\,\mu$s). Solid lines are fits using Eq.~\eqref{eq:empirical}. (b) Muon-spin asymmetry $A_{\rm LF}(t)$ measured under different LF at $T\simeq0.1$\,K. Left panel: enlarged short-time scale; right panel: long-time tail. The dashed lines represent simulated depolarization curves using Eq.~\eqref{eq:composite}. Solid lines are the fits using Eq.~\eqref{eq:sporadic}. (c) Temperature dependence of the ZF muon-spin relaxation rates ($\lambda_1$ and $\lambda_2$). Inset: temperature dependence of the stretching exponent $\beta$. (d) Temperature dependence of the fractional weight $\alpha$. Inset: LF dependence of the contribution $\phi$ from \emph{sporadic field fluctuations}.}
\end{figure*}

Next, we turn our attention towards $\mu$SR measurements. $\mu$SR probes spin dynamics over a frequency range complementary to NMR. 
Further, unlike NMR, $\mu$SR can be performed in true zero-field, eliminating any external perturbation of the intrinsic ground state.

Figure~\ref{Fig3}(a) presents the zero-field (ZF) muon-spin asymmetry spectra, $A_{\rm ZF}(t)$, measured on polycrystalline KBFPO in the $T$-range 30~mK $\leq T\leq$ 160~K. For $T > 10$~K, the relaxation follows a simple exponential decay. Upon cooling below 10~K, an additional fast initial depolarization appears, becoming increasingly Gaussian-like at lower temperatures. The continuous, monotonic loss of polarization without any coherent oscillations demonstrates the absence of static LRO~\cite{Amato2024}. Moreover, we did not observe any sign of 1/3-tail, expected for static, randomly oriented moments~\cite{Uemura546,Keren054403}. This suggests that the spins remain dynamic rather than static and/or frozen glassy state.

To describe the relaxation below 10~K, we tested several non-oscillatory functions, including single stretched exponentials and sums of exponentials. None provided a satisfactory description over the full time and temperature range. Instead, the spectra could be consistently modeled by the empirical function,
\begin{equation}
	\label{eq:empirical}
	A_{\rm ZF}(t) = A_0\bigl[\alpha e^{-(\lambda_1 t)^\beta} + (1 - \alpha)e^{-\lambda_2t}\bigr] + A_{\rm b}G_{\rm KT},
\end{equation}
where $\alpha$ and $(1 -\alpha)$ quantify the relative weight fractions of the two relaxation channels experienced by the sample. Here, $\lambda_1$ and $\lambda_2$ are relaxation rates associated with the fast early-time decay ($t\le0.3~\mu$s) and the slower long-time component ($t\ge0.3~\mu$s), respectively. The $A_0 = 0.241(1)$ and $A_{\rm b} = 0.005(1)$ are the contributions from muons stopping sites at the sample and Cu sample holder, respectively. The background relaxation is described by the static Gaussian Kubo-Toyabe function, $G_{\rm KT} = \tfrac{1}{3} + \tfrac{2}{3}\bigl(1 - \sigma^2 t^2\bigr)\exp\bigl[-\tfrac{1}{2}\sigma^2t^2\bigr]$, with $\sigma = 0.38~\mu$s$^{-1}$ fixed at the typical value expected for Cu~\cite{Kadono23} [see Fig.~S12(b) in SM]~\cite{SM}. Figures~\ref{Fig3}(c) and (d) show the extracted parameters ($\lambda_1$, $\beta$, $\lambda_2$, and $\alpha$) as a function of temperature.

At higher temperatures, the relaxation is purely exponential (rather than root‑exponential, as one expect in dilute systems), consistent with a dense network of fluctuating Fe$^{3+}$ ($S=5/2$) moments. Using the estimated nearest-neighbor exchange coupling $J/k_{\mathrm{B}}\simeq2$~K (from DFT calculations) and number of nearest neighbors $z = 6$, the high-temperature fluctuation rate can be approximated as $\nu=\sqrt{z}JS/\hbar \approx 1.6\times10^{6}~\mu$s$^{-1}$~\cite{Uemura3306}. In this dense magnetic environment, the fluctuating field distribution at the muon site can be approximated as Gaussian, with a width $\Delta = \gamma_\mu\,\sqrt{\langle(\mu_{0} H_{\mathrm{loc}} - \langle \mu_{0} H_{\mathrm{loc}}\rangle)^2\rangle}$, where $\gamma_\mu/2\pi=135.53$~MHz/T. Above 40~K, $\lambda_2$ is nearly temperature independent ($\sim 0.14~\mu$s$^{-1}$). Using $\lambda \approx0.14~\mu$s$^{-1}$ at ZF ($H_\mathrm{LF} = 0$) in the Redfield formula~\cite{Redfield1965}, $\lambda = 2\Delta^2\nu / \bigl[\nu^2 + \bigl(\gamma_\mu \mu_{0}H_{\mathrm{LF}}\bigr)^2\bigr]$ yields $\Delta\approx 335~\mu$s$^{-1}$. This confirms that the system remains in the fast fluctuating regime ($\nu\gg\Delta$).

As the temperature falls below $\sim 20$~K, $\lambda_2$ increases gradually, signaling a progressive slowing down of fluctuating moments, similar to $1/T_1$. Cooling further below 10~K, an additional faster relaxation component $\lambda_1$ appears. Both $\lambda_1$ and $\lambda_2$ eventually saturate below $T^*\simeq3.5$~K, consistent with a dynamic ground state and reminiscent of the behavior observed in QSL candidates~\cite{Khuntia107203,Ivica157204}. Moreover, the fractional weight $\alpha$ associated with the fast component, as well as the stretching exponent $\beta$, also shows a rapid increase and saturates at $\sim 0.79$ and to a purely phenomenological value of $\sim 3.8$ respectively at low temperatures. Constraining $\beta = 2$, the fit yields poor agreement with the early-time data [see Fig.~S12(a) in SM]~\cite{SM}. The transformation of the asymmetry shape below $T^*$ from exponential to Gaussian-like indicates that the spin fluctuations have slowed down significantly on approaching a regime where $\nu$ becomes comparable to $\Delta$.

Further evidence for persistent spin dynamics comes from the longitudinal-field (LF) dependent relaxation data at $T\simeq0.1$~K [Fig.~\ref{Fig3}(b)]. To evaluate these fluctuations, we analyzed the data using a composite function~\cite{Cho014439,Khuntia107203}
\begin{equation}
	\label{eq:composite}
	A_{\rm LF}(t) = A_{0}G_{\mathrm{DKT}}\bigl(t,\Delta,\nu,\mu_{0}H_{\mathrm{LF}}\bigr)
	+ A_{\rm b}G_{\mathrm{KT}}\bigl(t,\Delta_{\rm b},\mu_{0}H_{\mathrm{LF}}\bigr).
\end{equation}
Here, the first term describes the sample relaxation and decoupling behavior following a Dynamic Gaussian Kubo-Toyabe (DKT) function~\cite{Hayano850,Keren10039} under applied field, while the second term captures the contribution from muons stopping in the holder, which decouples rapidly in a small field of $\mu_0 H_{\mathrm{LF}}\simeq0.01$~T. This model gives a reasonable fit to the zero-field data, yielding $\Delta\approx16.1~\mu$s$^{-1}$ and $\nu\approx13.9~\mu$s$^{-1}$, suggesting that the spins approach a quasi-static limit ($\nu\leq\Delta$) at low temperature. However, applying these parameters to simulate field dependence exposes a clear discrepancy: in a purely quasi-static scenario, a field of about $\sim 0.1$~T would strongly suppress the relaxation, and 0.5~T should decouple it entirely. In contrast, even 1.5~T only partially decouples the relaxation [Fig.~\ref{Fig3}(b)], demonstrating that a simple quasi-static picture is inadequate and instead points to the presence of much stronger spin dynamics.

This led us to adopt a dynamical model based on \emph{sporadic field fluctuations}, which was introduced to explain persistent Gaussian relaxation in kagome bilayer chromates~\cite{Uemura3306,Bono187201}. In this framework, the relaxation function uses the same DKT form scaled by a factor $f$, namely $G_{\mathrm{DKT}}\bigl(t,f,\Delta,f\nu,fH_{\mathrm{LF}}\bigr)$. Here, the relaxation in the presumed SL state has been explained to arise from the deconfined spinon excitations that intermittently pass near the muon site, producing transient local magnetic fields during only a fraction $ft$ of the muon lifetime. For the remaining time interval $(1 - f)t$, the spins form a nearly non-magnetic, possibly singlet background that contributes negligible relaxation.

To describe the field dependence over the entire range, we replaced the first term in Eq.~\eqref{eq:composite} with the expression proposed by Bono et al~\cite{Bono187201},
\begin{equation}
	\label{eq:sporadic}
	\begin{aligned}
		A_{\rm LF}(t) =\; & A_{0}\bigl[
		\phi G_{\mathrm{DKT}}^{sp}\bigl(
		t,f\Delta,f\nu,f\mu_{0}H_{\mathrm{LF}}
		\bigr)
		+(1-\phi)e^{-\lambda't}
		\bigr] \\
		& +A_{\rm b}G_{\mathrm{KT}}\bigl(
		t,\Delta_{\rm b},\mu_{0}H_{\mathrm{LF}}
		\bigr).
	\end{aligned}
\end{equation}
The first term describes the dynamic relaxation arising from sporadic fluctuations. Here, $\phi$ quantifies the fraction of muons experiencing this process, while $(1 - \phi)$ accounts for a conventional exponential relaxation originating from a minority component in which the spin dynamics are slower and more continuous. In analogy with the zero-field model given by Eq.~\eqref{eq:empirical}, the parameter $\phi$ plays the same role as $\alpha$, representing the relative weight of the dominant dynamic contribution. This model provides a good fit to the data [Fig.~\ref{Fig3}(b)], yielding $\nu=560(5)~\mu\mathrm{s}^{-1}$, $\Delta=555.5(1)~\mu\mathrm{s}^{-1}\approx\gamma_\mu \times 0.65$~T, and $f=0.028(1)$. Despite strong local fluctuations with rate $\nu$, the depolarization can appear Gaussian-like because the fluctuation rate satisfies $\nu\approx\Delta$~\cite{Bono187201,Cho014439}. At low fields, $\phi$ is close to unity, making it effectively redundant as a fitting parameter in this regime. As the applied field increases, $\phi$ decreases [inset of Fig.~\ref{Fig3}(d)], indicating a gradual crossover from predominantly intermittent dynamics to more conventional Markovian fluctuations and enabling us to track how $\lambda^{\prime}$ varies with field [see Fig.~S12(c) in SM]~\cite{SM}. It should be noted that the fits to the low-field data are not perfect, which may lead to some overestimation of $\phi$.

Nevertheless, incorporating this second component ($1-\phi$) emphasizes that not all regions of the sample contribute equally to the dynamic fluctuations, highlighting a degree of spatial inhomogeneity within the magnetic ground state. Analysis of the zero-field data leads to a similar conclusion: the ground state consists of a dominant dynamic phase (with $\alpha\approx0.79$) characterized by sporadic spin fluctuations and a minor component ($1-\alpha\approx0.21$) exhibiting slower and more heterogeneous relaxation. Such coexisting inhomogeneous dynamics have been observed in other highly frustrated magnets~\cite{Bono187201,Cho014439,Dey024407}. 
In this scenario, the observed robust dynamic with pronounced heterogeneity in a coupled trillium lattice, KBFPO, represents a remarkable finding, demonstrating the rich landscape of magnetic correlations in this compound.

Though the first magnetic study on KBFPO was reported nearly four decades ago, its magnetic ground state still remains elusive. Our findings pinpoint a dynamic ground state, which is in sharp contrast to an $L$-type ferrimagnetic transition near 4~K proposed by Battle \textit{et al.}~\cite{Battle16}. In KBFPO, Fe$^{3+}$ ions form a frustrated bipartite double-trillium lattice with strong AFM interactions. While a weak feature appears at $T^{*}\simeq3.5$~K in low-field $\chi(T)$, a conventional magnetic LRO or spin freezing is ruled out down to 30~mK. Below $T^{*}$, both local probes (NMR and $\mu$SR) reveal a spatially inhomogeneous dynamic state with two relaxation channels: a dominant and fast-relaxing component ($\sim 80$\%) coexisting with a slower, yet dynamic minority component ($\sim 20$\%). Moreover, the LF $\mu$SR data are well captured by a \emph{sporadic-fluctuation model}, drawing a parallel with the kagome bilayers, where a fluctuating RVB background coexists with slowly fluctuating spins~\cite{Uemura3306,Bono187201}. This inhomogeneity in spin dynamics appears to be a unique characteristic feature and a consequence of strong magnetic frustration due to trillium geometry in 3D.
Specific heat exhibits a broad maximum near $T^{*}$, which is field-independent, reminiscent of short-range spin correlations and contradicting a conventional magnetic LRO. A further evidence for the dynamic state is obtained from the quadratic temperature dependence of $C_{\rm mag}(T)$ at low-temperatures.

Altogether, KBFPO realizes a rare high-spin, 3D frustrated magnet that evades LRO and develops a \emph{mosaic-like} landscape of fluctuating spins with different dynamics. These findings, observed for the first time in a double trillium lattice, establish KBFPO as a unique material candidate and an ideal model system for exploring frustration-driven, dynamically inhomogeneous magnetism in 3D. Our findings in this Letter further call upon further experimental investigations, such as diffuse neutron scattering, chemical substitution, etc, to uncover the microscopic origin of the observed spin dynamics and to assess the potential proximity of KBFPO to a canonical SL phase.

\acknowledgments
SJS acknowledges Fulbright-Nehru Doctoral Research Fellowship Award No.~2997/FNDR/2024-2025 and the Prime Minister's Research Fellowship (PMRF) scheme, Government of India to be a visiting research scholar at the Ames National Laboratory. SJS and RN also acknowledges SERB, India, for financial support bearing sanction Grant No.~CRG/2022/000997 and DST-FIST with Grant No.~SR/FST/PS-II/2018/54(C). The research was supported by the U.S. Department of Energy, Office of Basic Energy Sciences, Division of Materials Sciences and Engineering. Ames National Laboratory is operated for the US Department of Energy by Iowa State University under Contract No. DEAC02-07CH11358. The work in Leipzig was funded by funded by the Deutsche Forschungsgemeinschaft (DFG, German Research Foundation) -- TRR 360 -- 492547816 (subproject B3). This work is partially based on experiments performed at the Swiss Muon Source S$\mu$S, Paul Scherrer Institute, Villigen, Switzerland.


%

\end{document}